\documentclass[letter]{aa}  
\usepackage{graphicx}
\usepackage{txfonts}
\begin{document}

   \title{Excitation of decay-less transverse oscillations of coronal loops by random motions}

   \author{A.N. Afanasyev
          \inst{1,2}
          \and
          T. Van Doorsselaere
          \inst{1}
          \and
          V.M. Nakariakov
          \inst{3,4}}
          
   \institute{Centre for mathematical Plasma Astrophysics, KU Leuven,
Celestijnenlaan 200B, box 2400, B-3001 Leuven, Belgium\\
              \email{afa@iszf.irk.ru}
        \and
             Institute of Solar-Terrestrial Physics of SB RAS, Irkutsk, Russia
        \and
            Centre for Fusion, Space and Astrophysics, University of Warwick, Coventry, UK
        \and
            St. Petersburg Branch, Special Astrophysical Observatory, Russian Academy of Sciences, St. Petersburg, Russia}

   \date{Received; accepted }

 
  \abstract
   {The relatively large-amplitude decaying regime of transverse oscillations of coronal loops has been known for two decades and interpreted in terms of magnetohydrodynamic kink modes of cylindrical plasma waveguides. In this regime oscillations decay in several cycles. Recent observational analysis has revealed the so-called decay-less small-amplitude oscillations, with a multi-harmonic structure being detected. Several models have been proposed to explain them. In particular, decay-less oscillations have been described in terms of standing kink waves driven with continuous monoperiodic motions of loop footpoints, in terms of a simple oscillator model of forced oscillations due to harmonic external force, and as a self-oscillatory process due to the interaction of a loop with quasi-steady flows. However, an alternative mechanism is needed to explain the simultaneous excitation of several longitudinal harmonics of the oscillation.}
   {We study the mechanism of random excitation of decay-less transverse oscillations of coronal loops.}
   {With a spatially one-dimensional and time-dependent analytical model taking into account effects of the wave damping and kink speed variation along the loop, we consider transverse loop oscillations driven by random motions of  footpoints. The footpoint motions are modelled by broad-band coloured noise.}
   {We have found the excitation of loop eigenmodes and analysed their frequency ratios as well as the spatial structure of the oscillations along the loop. The obtained results successfully reproduce the observed properties of decay-less oscillations. In particular, excitation of eigenmodes of a loop as a resonator can explain the observed quasi-monochromatic nature of decay-less oscillations and generation of multiple harmonics detected recently.}
   {We propose  the  mechanism  that  can interpret decay-less transverse  oscillations of coronal loops in terms of kink waves randomly driven at the loop footpoints.}

\titlerunning{Excitation of Decay-less Loop Oscillations by Random Motions}
    \authorrunning{Afanasyev et al.}
   \maketitle
%

\section{Introduction}

Observations of the Sun in the extreme-ultraviolet spectral band show that the solar corona is disturbed by numerous wave and oscillatory plasma motions  (e.g., see reviews by \citealp{Liu2014, Nakariakov2016SSRv..200...75N}). In particular, transverse oscillations of coronal loops have been detected and interpreted in terms of magnetohydrodynamic (MHD) kink modes of cylindrical plasma waveguides \citep{Aschwanden1999ApJ...520..880A, Nakariakov1999Sci...285..862N, Goddard2016A&A...585A.137G}. Analysis of observational data has revealed two different classes of transverse oscillations of coronal loops. The first class includes impulsively excited oscillations of relatively large amplitude that decay rapidly in several oscillation cycles \citep{Aschwanden1999ApJ...520..880A,Nakariakov1999Sci...285..862N}. The other class includes long-existing small-amplitude oscillations showing no significant damping \citep{Wang2012-L27, Tian_2012, Nistico2013, Nistico2014, Anfinogentov2015}. The displacements of coronal loops oscillating in this decay-less regime are usually less than the minor radius of the oscillating loop, and their periods scale with the loop length \citep{Anfinogentov2015}, whereas the loop velocity amplitude does not show any dependence on the loop length and period. Although the amplitude of oscillations is found to experience some variation at longer time scales, they show periodic and almost harmonic nature. Recently, \citet{Duckenfield2018ApJ...854L...5D} have reported on the first detection of multiple harmonics in decay-less oscillations. Such decay-less transverse oscillations are now in the focus of both theoretical and observational studies due to their seismological potential and possible role in wave plasma heating (e.g., \citealp{Stepanov_book_2012, Arregui2015}).

\begin{figure*}
\centering
\includegraphics[width=0.4\textwidth,clip=0.0]{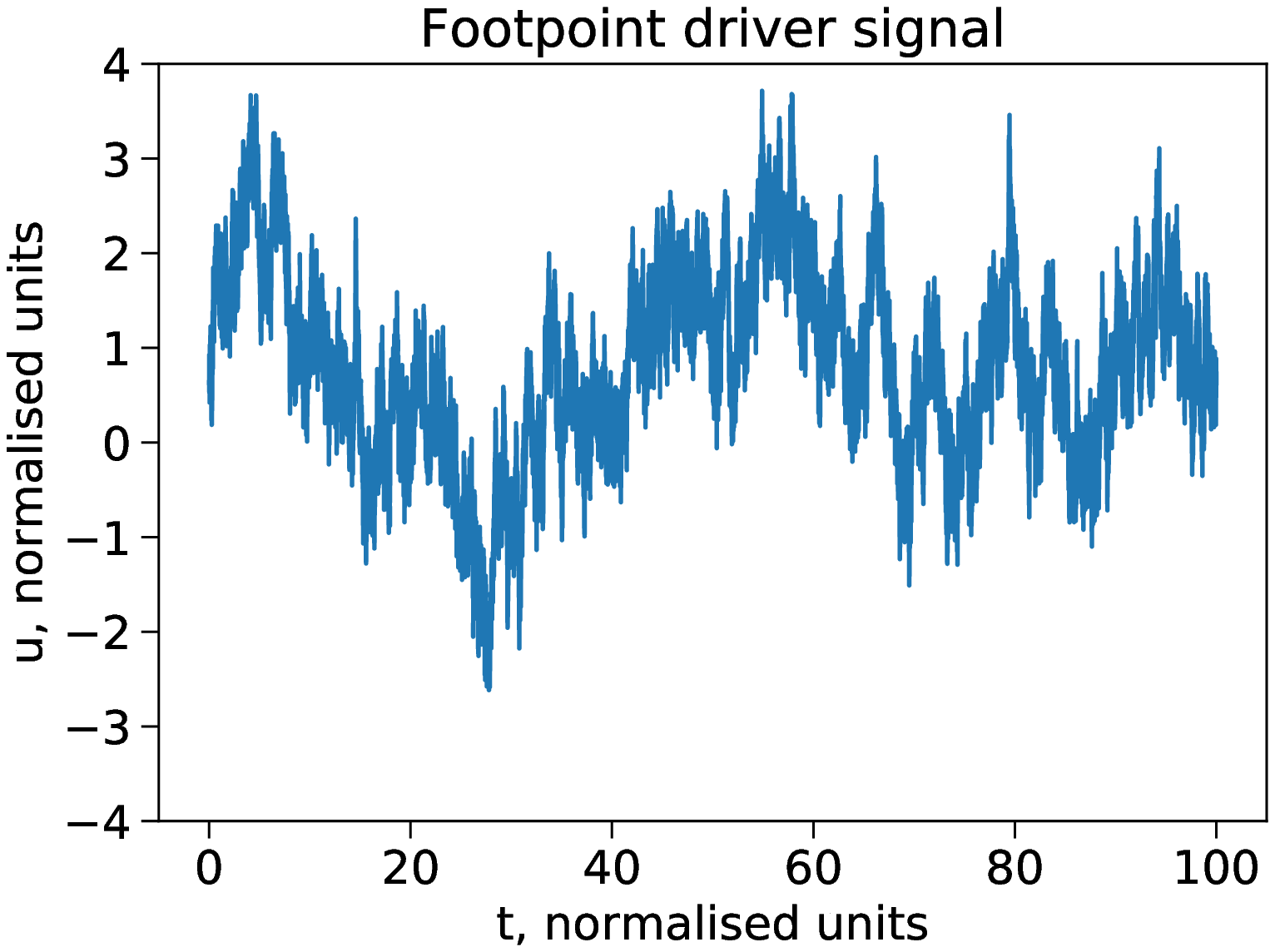}
\hspace{1cm}
\includegraphics[width=0.4\textwidth,clip=0.0]{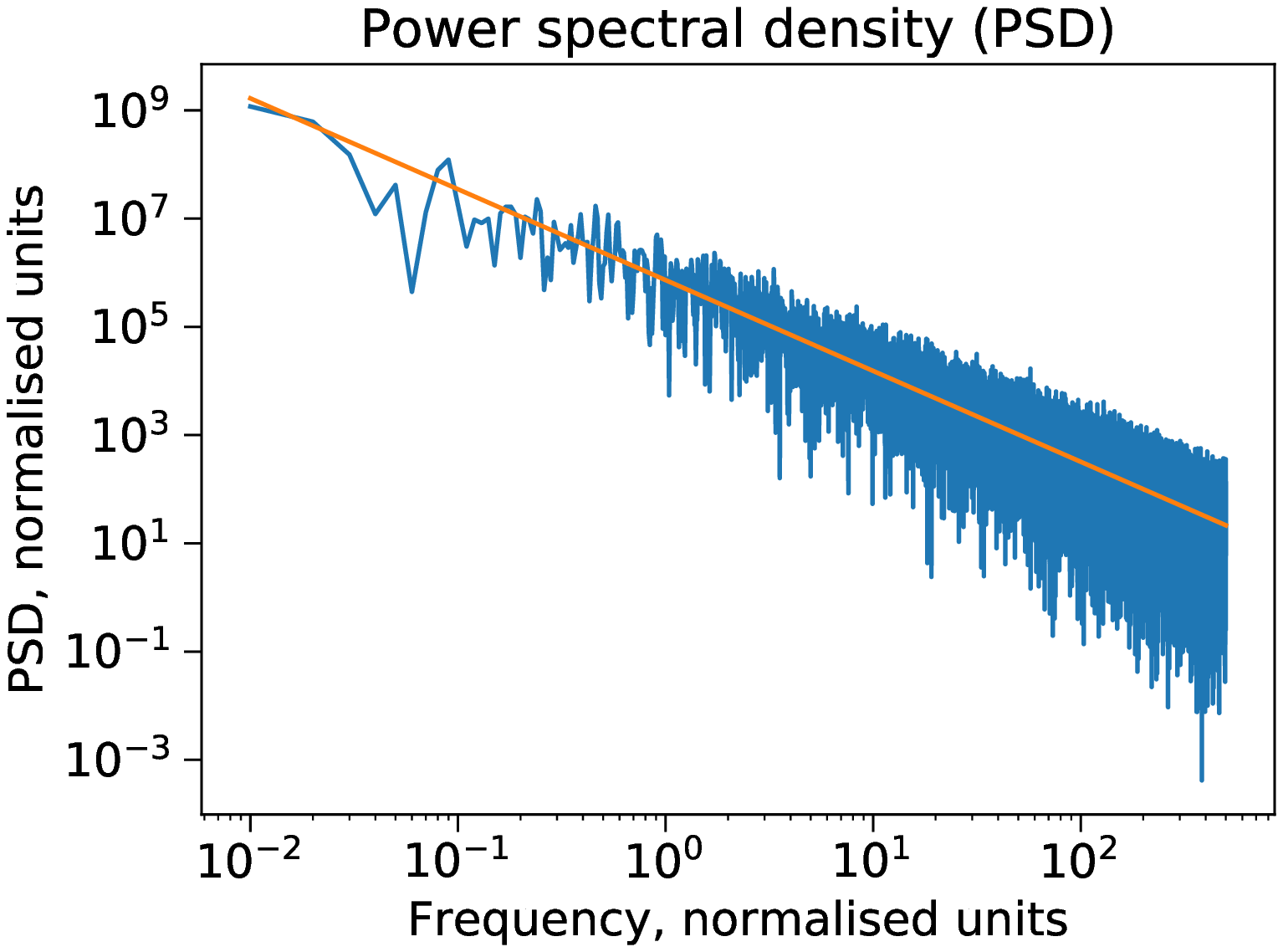}
\caption{Left loop footpoint displacement (left panel) and its Fourier spectrum (right panel). The orange line shows the spectral power-law fall-off with index $\delta = 1.66$.}
\label{Fig-input-signal-and-spectrum}%
\end{figure*}

A number of efforts were made to explain the nature of transverse oscillations in the decay-less regime. Obviously, energy should be continuously supplied to the system to maintain oscillations with no visible damping. First, a simple analytical model of damped linear oscillator with an external continuously driving force was proposed to interpret the observed oscillations \citep{Nistico2013, Nistico2014}. Mathematically, the model was represented with a non-homogeneous ordinary differential equation of Helmholtz type. \citet{Nistico2013} considered decay-less oscillations in terms of forced oscillations due to a non-resonant harmonic external force. \citet{Nistico2014} discussed them as a result of a random external force in the presence of significant damping possibly due to dissipation and resonant absorption, explaining the observed intermittent variation of the oscillation amplitude and phase, which was observed specifically in the case under their analysis. \citet{Nakariakov2016A&A...591L...5N} also discussed the random nature of driving force, however, concluded that resulting oscillations experienced highly intermittent variation of the oscillation amplitude and phase.

A more sophisticated nonlinear model was proposed by \citet{Nakariakov2016A&A...591L...5N}, in which the authors interpreted decay-less transverse oscillations as a self-oscillatory process due to the interaction of loops with quasi-steady flows at the loop footpoints. The mathematical formalism included a nonlinear non-homogeneous term, so the governing equation was reduced to the Rayleigh oscillator equation with a constant flow speed as a parameter. Its solution corresponding to the limit cycle solution could explain the sustained nature of observed oscillations, similarly to oscillations of a violin string forced with a bow.

However, the low-dimensional approach based on the oscillator or self-oscillator models have several important shortcomings. In an oscillating loop the source of energy is likely to be spatially non-uniform, and, mostly likely, acts at the loop footpoints. This fact cannot be adequately described by a low-dimensional model which accounts for the time dependence only. In particular, an important missing effect is the simultaneous excitation of several longitudinal harmonics of the oscillation.

Along with analytical investigations, a number of numerical efforts were made to analyse the nature of transverse oscillations in terms of standing kink waves driven continuously at the loop footpoints, using three-dimensinal MHD simulations (e.g., \citealp{Karampelas2017, Karampelas2019A&A...623A..53K, Howson2017, Guo2019}, see also \citealp{ofman1996ApJ...456L.123O}). More attention in those studies was paid to the development of the Kelvin-Helmholtz instability (KHI) and coronal plasma heating. \citet{Afanasyev2019} studied the response of a coronal loop to continuous monoperiodic drivers of different frequencies and found its resonant behaviour. \citet{Karampelas-2019frontiers} obtained that the amplitude of decay-less oscillations weakly depends on the driver strength. \citet{antolin2016} studied the connection between decay-less oscillations and line-of-sight effects due to the formation of KHI vortices. 

In this Letter we demonstrate a mechanism for the random driving of decay-less oscillations at the loop footpoints within a spatially one-dimensional and time-dependent analytical model, which allows one to interpret decay-less loop oscillations in a way consistent with recent observational findings.

\begin{figure*}
\centering
\includegraphics[width=0.4\textwidth,clip=0.0]{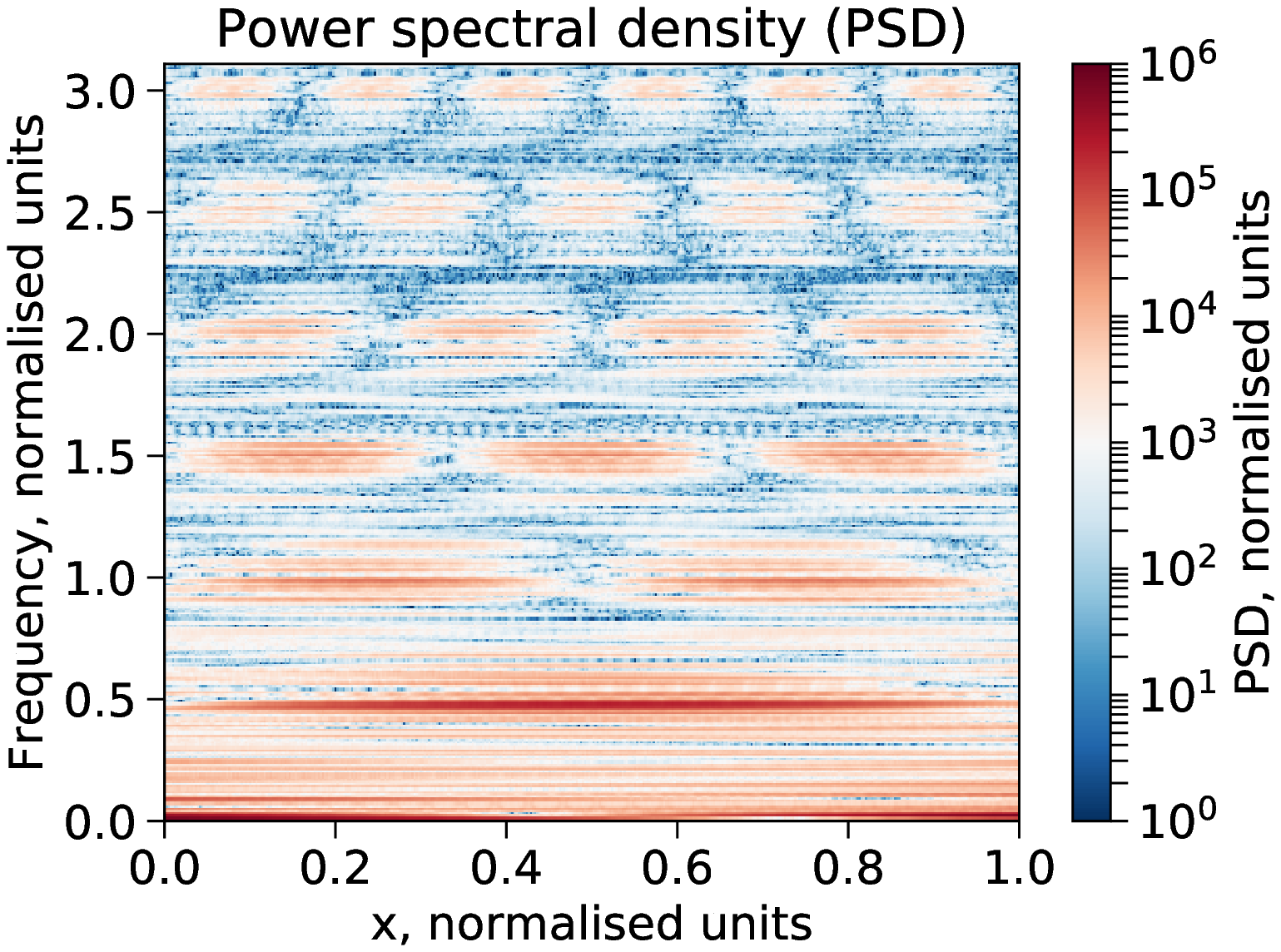}
\hspace{1cm}
\includegraphics[width=0.4\textwidth,clip=0.0]{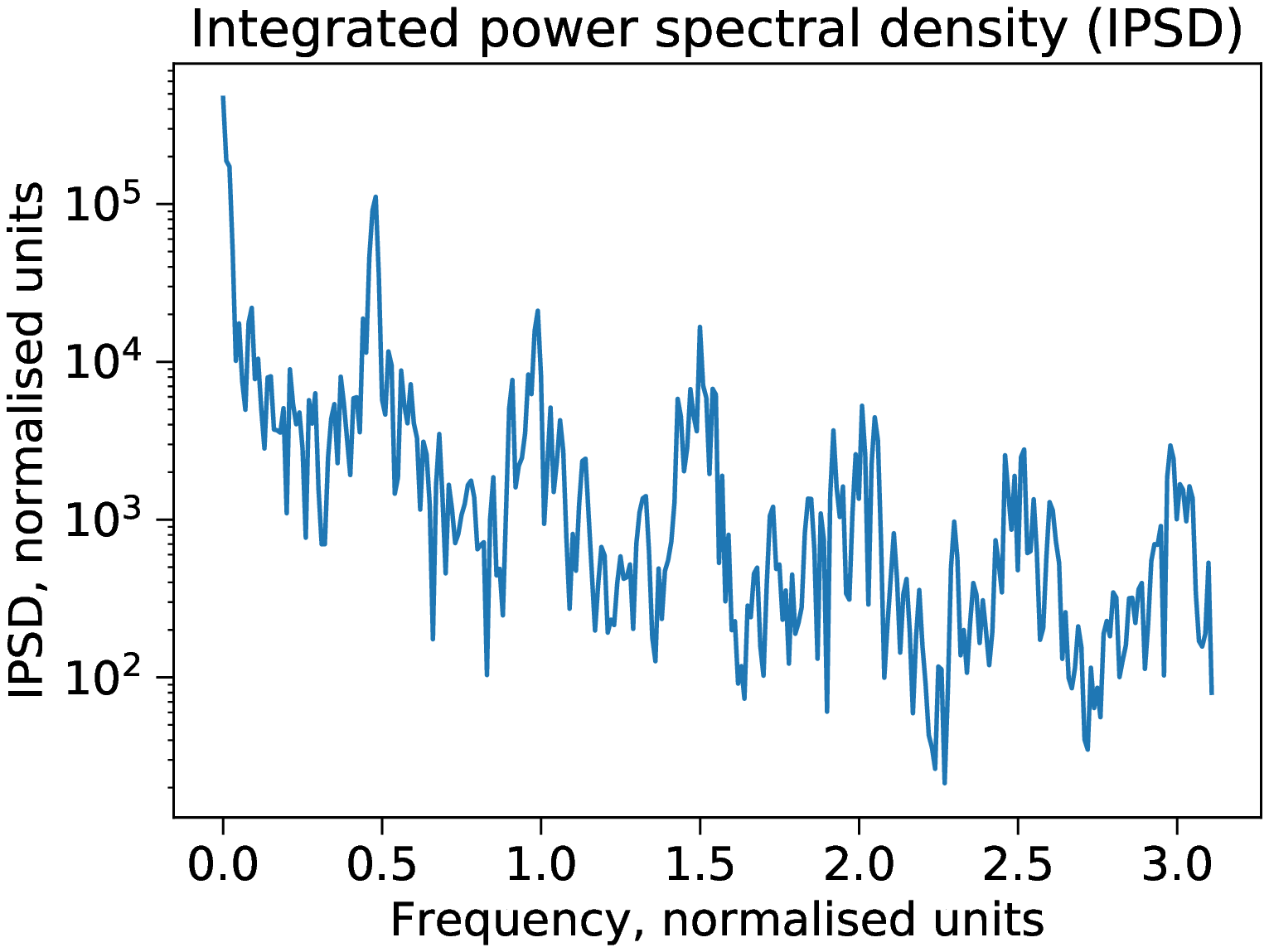}
\caption{Power spectral density (left) and power spectral density integrated over $x$ (right) of displacements of a coronal loop perturbed at its footpoints with a power-law random noise. The case of the non-stratified plasma with a constant kink speed is shown.}
\label{Fig-spectrum-damping-constspeed}%
\end{figure*}
\begin{figure*}
\centering
\includegraphics[width=0.4\textwidth,clip=0.0]{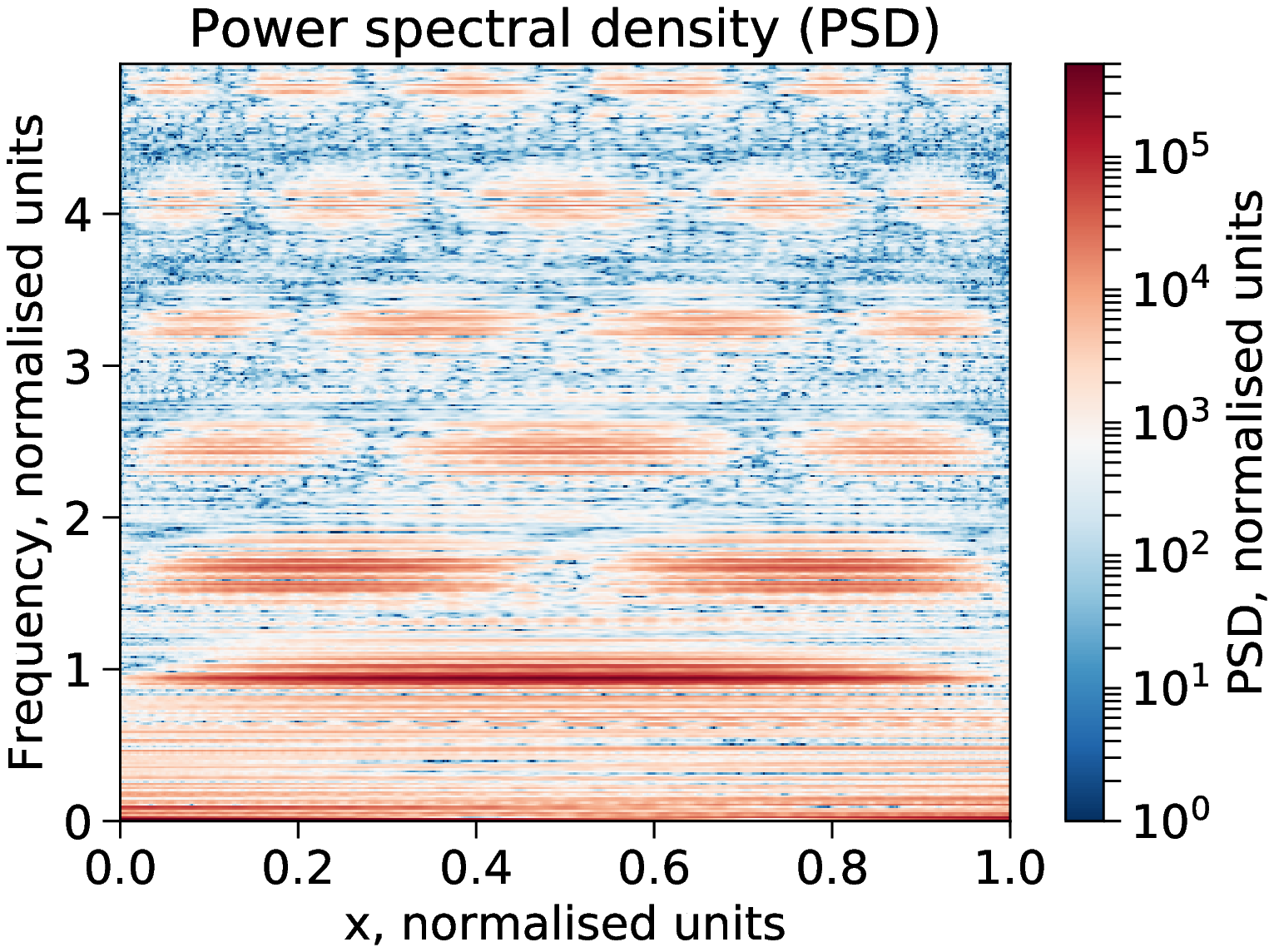}
\hspace{1cm}
\includegraphics[width=0.4\textwidth,clip=0.0]{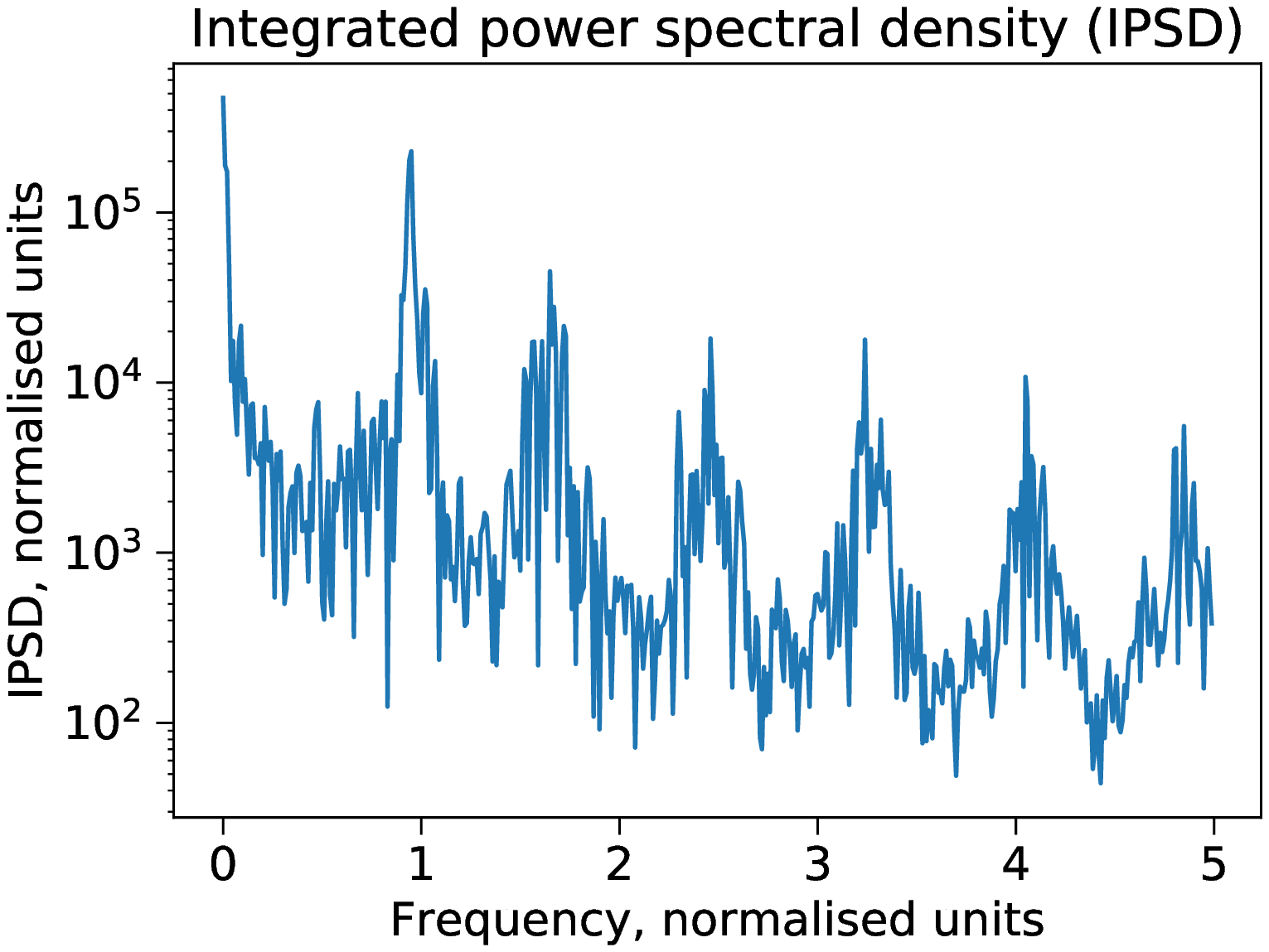}
\caption{Same as in Figure~\ref{Fig-spectrum-damping-constspeed}. The case of the stratified plasma with a kink speed varying along the loop is shown.}
\label{Fig-spectrum-damping-varspeed}%
\end{figure*}

\section{Model}
\label{sec:model}

We consider transverse coronal loop oscillations of small amplitude as standing kink waves in a magnetic cylinder. \citet{Dymova2005} showed that linear non-axisymmetric wave dynamics in a loop filled in with a low-$\beta$ plasma can be described in the long-wavelength limit with a one-dimensional wave equation. A similar approach based on the ordinary differential (Helmholtz) equation of oscillations was used by \citet{Verth-etal-2007} and \citet{ErdelyiVerth2007} who studied frequency shifts and amplitude variations in the oscillation eigenmodes due to the non-uniformity along the field. Following those studies, we model decay-less oscillations as transverse oscillations of an elastic string. The background magnetic field is assumed constant and strong enough, so that the low-$\beta$ plasma condition is fulfilled. On the other hand, we take into account the plasma stratification due to the gravity varying with height owing to the loop curvature. So, we consider the governing equation for transverse displacements of the string, $u$, as a function of distance $x$ along the string in the following form
\begin{equation}
    \frac{\partial^2 u} {\partial t^2} + \alpha  \frac{\partial u} {\partial t} = C_\mathrm{k}^2\left( x \right) \frac{\partial^2 u} {\partial x^2},
\label{Eq-wave-equation}
\end{equation}
where the wave damping characterised by a constant coefficient $\alpha$ is introduced. The nature of the damping is resonant absorption of kink waves in radially non-uniform plasma cylinders \citep{ruderman2002,goossens2002} and possibly effects of the KHI development (e.g., \citealp{heyvaerts1983A&A...117..220H, Terradas2008, Karampelas2017}). We have chosen $\alpha = 0.5$ (in normalised units) based on the results of our numerical analysis \citep{Afanasyev2019}, using the ratio $\Delta f /f$ as a criterion. The varying kink speed, $C_\mathrm{k} \left( x \right)$, is determined by density decrease with height as
\begin{equation}
   C_\mathrm{k}\left( x \right) =  C_0 \exp \left( \frac {L}{2 \pi H} \sin \left( \frac{\pi x}{L} \right) \right), 
\label{Eq-sound-speed}
\end{equation}
where $C_0$ is the value of the kink speed at the loop footpoints, $L \approx 200$~Mm is the length of the loop, $H \approx 40$~Mm is the density scale height determined by a plasma temperature of $0.8$~MK.

We normalise the equation quantities to the loop length $L$ and time $T = L/C_0$, which is the wave transit time in the case of a constant kink speed $C_0$. Initially, the loop is at rest. The boundary conditions include random drivers at both footpoints, mimicking random footpoint buffeting of coronal loops. The input driver signals are so-called coloured noise, so the power spectral density $S$ of the random footpoint displacements has a power-law fall-off, $S \propto f^{-\delta}$, where $f$ is the frequency, $\delta > 1$. For instance, we assume a spectral index $\delta = 1.66$ so that more energy is contained in the low-frequency range. Random motions driving the photospheric footpoints of magnetic tubes were also used in models of the Alfv\'en turbulence \citep{van_Ballegooijen_2011, Asgari_Targhi_2012, Asgari_Targhi_2013}. On the other hand, the use of such turbulent power-law spectra is supported by observations of the dynamics of magnetic bright points (e.g., \citealp{Abramenko_2011, Chitta_2012}), which might be regarded as the motion of loop footpoints, as well as by observations of sunspot oscillations \citep{Kolotkov_2016}. The noise signals have been generated using Python package {\sl colorednoise~1.1.0} implementing the algorithm by \citet{Timmer1995A&A...300..707T} to generate Gaussian distributed noise with a power-law spectrum.  Figure~\ref{Fig-input-signal-and-spectrum} shows the input noise signal at the left footpoint and its power spectrum.

We solve Eqs.~\ref{Eq-wave-equation} and \ref{Eq-sound-speed} numerically, using central differences of second order accuracy in time and space for derivatives. We calculate the loop dynamics up to $100 \, T$.

\section{Results and discussion}
\label{sec:results}

The loop experiences random displacements of its axis, responding to the excitation of the footpoints with coloured noise. To analyse the structure of the displacements, we perform Fourier analysis. We apply the fast Fourier transform only with respect to the temporal part of the two-dimensional output signal $u \left( x, t \right)$. Thus, for each value of the spatial coordinate $x$, we have a one-dimensional Fourier spectrum. Such an approach allows us to identify directly the spatial structure of the harmonics detected.

Figure~\ref{Fig-spectrum-damping-constspeed} demonstrates the $x-f$~diagram (left panel) in the case of the non-stratified plasma ($H \to  \infty$) with a constant kink speed $C_0$. The diagram shows that random loop oscillations have the explicit harmonic structure in time and the spatial structure corresponding to standing wave modes as observed in the decay-less regime. The maxima of the power spectral density (PSD) are located at positions of the standing wave anti-nodes of transverse displacements. The enhanced values of the PSD near zero frequency appear to be due to the power-law spectral fall-off of the input signals. The obtained pattern shows the excitation of eigenmodes of the loop as a resonator, extending the results of \citet{Afanasyev2019}. The right panel of Fig.~\ref{Fig-spectrum-damping-constspeed} shows the PSD integrated over the spatial coordinate $x$, demonstrating the frequency structure only. Once again, we stress the similarity with Fig.~2 of \citet{Afanasyev2019}. Also, the simulations performed for different values of the damping coefficient $\alpha$ show that damping determines the width and height of the spectral peaks as expected from the general theory of oscillations.

Figure~\ref{Fig-spectrum-damping-varspeed} shows the same type of diagram in the case of a varying kink speed $C_\mathrm{k} \left( x \right)$ described by Eq.~(\ref{Eq-sound-speed}). Positions of the maxima of the PSD are shifted, so the ratios of the frequencies of harmonics are no longer equal to integer numbers. The shift is due to the gravitational stratification of the plasma, and in particular, because of the higher kink speed at the loop apex (see also \citealp{Andries2005}). The ratio of the frequencies of the second and first harmonics is about $1.7$, as seen in Figure~\ref{Fig-spectrum-damping-varspeed}. Recently, \citet{Duckenfield2018ApJ...854L...5D} detected the frequency ratio for decay-less transverse oscillations of a coronal loop to be $1.4$. It would be of some interest to develop a seismological technique allowing one to extract useful information on the loop (in particular, plasma temperature inside the loop) by using detected values of the harmonic ratio. Although a more sophisticated and rigorous model would be required for that (see, e.g., \citealp{Andries_2005_seismology}).

\begin{figure}
\centering
\includegraphics[width=0.4\textwidth,clip=0.0]{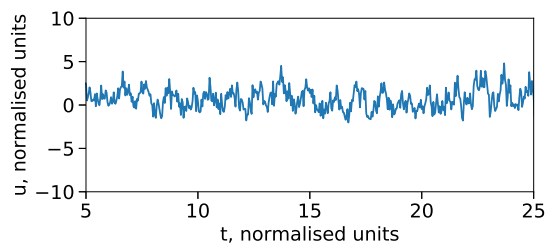}
\caption{Displacement of the apex of a coronal loop perturbed with a power-law random noise at the loop footpoints.}
\label{Fig-displacement-in-loop-apex}%
\end{figure}

Figure~\ref{Fig-displacement-in-loop-apex} shows the displacements of the loop axis at the loop apex. It is remarkable that in our model, under the random excitation of its footpoints the loop shows oscillations of harmonic nature with slowly varying amplitude over several oscillation cycles, as seen in observations of the loop dynamics in the extreme-ultraviolet band (e.g., see Fig.~1 in \citealp{Nakariakov2016A&A...591L...5N}, see also \citealp{Liu2014}). The oscillations obtained in our simulations contain a noise component leading to small-scale intermittent pattern, however, this component might be unresolved with Atmospheric Imaging Assembly on the Solar Dynamics Observatory (SDO/AIA) due to line-of-sight effects and resolution limitations. Indeed, the displacement amplitudes of decay-less oscillations are comparable with the resolution of the instrument (see e.g., \citealp{Nakariakov2016A&A...591L...5N, Anfinogentov2016}).

The proposed model successfully reproduces the observed properties of decay-less oscillations. The periods of the excited harmonics are determined by the length of the loop and kink speed, which is in agreement with detected scaling of periods of decay-less oscillations with loop lengths. On the other hand, the velocity amplitude of oscillations is entirely determined by the random noise at the loop footpoints (for the given plasma stratification and damping), so no scaling of oscillation amplitudes with loop lengths should be detected in this case. However, the scaling laws of observable parameters with the efficiency of the damping mechanism and the power of the driver in our model yet need to be established in follow-up studies. The excitation of eigenmodes of a loop as a resonator can explain the observed quasi-monochromatic nature of decay-less oscillations. The presence of damping in the solar corona most probably due to the resonant absorption of transverse waves is compensated by the energy supply at the loop footpoints, balancing and providing a sustained oscillation regime. In particular, the model explains the appearance of multiple harmonics, which was elusive in the previously proposed models. Taking into account the variation of the kink speed along the loop due the gravity stratification allows us to obtain a ratio of harmonic frequencies different from integer numbers, which is also in good agreement with recent observations.

\section{Summary and concluding remarks}
\label{sec:discussion-conclusion}

In this Letter we have proposed a mechanism that interprets decay-less transverse oscillations of coronal loops in terms of kink MHD waves randomly driven at the loop footpoints. The mathematical formalism used is based on a spatially one-dimensional and time-dependent analytical model including the wave equation with two-boundary excitation with a coloured random noise. Unlike the previous models, we have introduced the spatial dimension to take into account the fact that the energy supporting the decay-less regime of loop oscillations is supplied only at the loop footpoints. Another important feature of our model is random excitation of loop oscillations at the footpoints, which seems to be the most realistic in the solar atmosphere. In contrast with the self-oscillation model \citep{Nakariakov2016A&A...591L...5N}, which considers the energy supply by the plasma flows with characteristic time scales much longer than the oscillation period (e.g., associated with supergranulation), in the present model the energy supply is provided by footpoint motions with the characteristic time scales comparable to the oscillation periods (e.g., associated with granulation). The results obtained explain such observed properties of decay-less oscillations as a quasi-monochromaticity, period-length scaling, sustainability, and in particular, excitation of multiple harmonics, and specific values of ratios of the harmonic frequencies. 

Our simple model inevitably has a number of shortcomings. In particular, the governing equation (Eq.~\ref{Eq-wave-equation}) does not adequately describe the oscillation damping caused by the plasma non-uniformity across the field, i.e., resonant absorption. So, the damping is introduced in a way excluding its frequency dependence. We expect that more sophisticated analytical derivations would be able to allow the low-frequency eigenmodes to dominate even more significantly, which could lead to a more pronounced harmonic spatial structure of the loop oscillations. However, in follow-up work we will focus on three-dimensional numerical MHD simulations of decay-less loop oscillations. Thus, the main purpose of the analytical model presented in this Letter is to show the principal results valuable for the physical understanding of the problem.

\begin{acknowledgements}
We are thankful to Dr C.R. Goddard and T. Duckenfield for discussion of the idea of the work.
The work was supported by the European Research Council (ERC) under the European Union’s Horizon 2020 research and innovation programme (grant agreement No.~724326) and the C1 grant TRACEspace of Internal Funds KU Leuven. A.N.A. and V.M.N. acknowledge support from the Russian Foundation for Basic Research under grant 17--52--80064 BRICS-A. 
The results were inspired by discussions at the ISSI-Bern meeting ``Quasi-periodic Pulsations in Stellar Flares: a Tool for Studying the Solar-Stellar Connection'' and at the ISSI-Beijing meeting ``Pulsations in solar flares: matching observations and models''.
\end{acknowledgements}

\bibliographystyle{aa}
\bibliography{biblio_random_excitation}

\end{document}